\documentstyle[a4,11pt] {article}
\begin{document}

\setlength{\baselineskip}{20pt}

\noindent
{\Large\bf Comment on `Oblique Propagation and Dissipation of Alfv\'{e}n Waves 
in Coronal Holes' by Srivastava and Dwivedi}
\vspace{4mm}

\setlength{\baselineskip}{15pt}

\noindent
{\bf Suresh Chandra$^1$ and Lalan Prasad$^2$}

\noindent
$^1$ School of Physics, S.M.V.D. University, Katra 182 320, (J\&K) \\
$^2$ Dept. of Physics, M.B. Govt. P.G. College, Haldwani 263 241

\vspace{2mm}

\noindent
E-mail: sch@iucaa.ernet.in; lprasad@iucaa.ernet.in

\vspace{2mm}

\noindent
Paper has been submitted to J. Astrophy. Astron. on 2 June 2007 and decision
is awaited for quite long time.

\vspace{6mm}

 \noindent
{\bf Abstract:} We comment on the recent paper by Srivastava and Dwivedi (J.
Astrophs. Astr. {\bf 28}, 1, 2007). In the said paper the derived dispersion 
relation $\nu \eta \cos^4 \theta \ k^4 + [v_A^2 - i \omega (\nu + \eta)]
\cos^2 \theta k^2 - \omega^2 = 0$ seems to be in error. Consequently, the 
conclusion drawn in their paper are not reliable. 

\section{Introduction}

In their recent paper, Srivastava and Dwivedi (2007, henceforth SD) investigated
the oblique propagation and dissipation of Alfv\'{e}n waves in
coronal holes where the basic equations used by them are
\begin{eqnarray}
\rho \frac{\partial \hspace{-1mm} \stackrel{\rightarrow}{v}}{\partial t} + \rho 
(\stackrel{\rightarrow}{v}. \nabla)\stackrel{\rightarrow}{v} = \frac{1}{\mu}
(\nabla \times \stackrel{\rightarrow}{B}) \times \stackrel{\rightarrow}{B} + 
\rho \nu \nabla^2 \hspace{-1mm} \stackrel{\rightarrow}{v} \hspace{1cm}
\mbox{Momentum equation} \label{eq:1} \\
\frac{\partial \hspace{-1mm} \stackrel{\rightarrow}{B}}{\partial t} = \nabla 
\times (\stackrel{\rightarrow}{v} \times \stackrel{\rightarrow}{B}) + \eta 
\nabla^2 \hspace{-1mm} \stackrel{\rightarrow}{B} \hspace{3.8cm} 
\mbox{Induction equation} \label{eq:2} \\
\nabla . \stackrel{\rightarrow}{B} = 0 \hspace{5.9cm} \mbox{Magnetic flux 
conservation} \label{eq:3}
\end{eqnarray}

\noindent
where $\stackrel{\rightarrow}{v}$ is the velocity, $\stackrel{\rightarrow}{B}$ 
the magnetic field and $\rho$, $\mu$, $\eta$, $\nu$ are respectively the mass 
density, magnetic permeability, magnetic diffusivity and the coefficient of 
viscosity. For small perturbations (represented by a suffix 1) from the 
equilibrium (represented by a suffix 0) (Priest, 1982):
\begin{eqnarray}
\rho = \rho_0 + \rho_1 \hspace{2.5cm} \stackrel{\rightarrow}{v} = \stackrel{
\rightarrow}{v}_1 \hspace{2.5cm} \stackrel{\rightarrow}{B} = \stackrel{
\rightarrow }{B}_0 + \stackrel{\rightarrow}{B}_1  \nonumber
\end{eqnarray}

\noindent
and after linearization, equations (\ref{eq:1}) and  (\ref{eq:2}), reduces to 
 (Pek\"unl\"u et al, 2002; SD) 
\begin{eqnarray}
\rho_0 \frac{\partial \hspace{-1mm} \stackrel{\rightarrow}{v}_1}{\partial t} = 
\frac{1}{\mu} (\stackrel{\rightarrow}{B}_0 . \nabla) \stackrel{\rightarrow}{B}_1
+ \rho_0 \nu \nabla^2 \hspace{-1mm} \stackrel{\rightarrow}
{v}_1 \label{eq:4} \\
\frac{\partial \hspace{-1mm} \stackrel{\rightarrow}{B}_1}{\partial t} = 
(\stackrel{\rightarrow}{B}_0 . \nabla) \stackrel{\rightarrow}{v}_1 + \eta 
\nabla^2 \hspace{-1mm} \stackrel{\rightarrow}{B}_1 \hspace{0.9cm}\label{eq:5} 
\end{eqnarray}

\noindent
Here, the magnetic field $\stackrel{\rightarrow}{B}_0$ is taken uniform as well
as time independent. A usual practice has been to consider plane-wave solutions
(Priest, 1982; Porter et al., 1994; Pek\"unl\"u et al., 2002; Kumar et al., 
2006) 
\begin{eqnarray}
\stackrel{\rightarrow}{v}_1 = \stackrel{\rightarrow}{v} \ \mbox{e}^{i(\stackrel
{\rightarrow}{k} . \stackrel{\rightarrow}{r} - \omega t)} \hspace{3cm}
\stackrel{\rightarrow}{B}_1 = \stackrel{\rightarrow}{B} \ \mbox{e}^{i(\stackrel
 {\rightarrow}{k} . \stackrel{\rightarrow}{r} - \omega t)} \label{eq:8}
\end{eqnarray}

\noindent
where $\stackrel{\rightarrow}{k}$ is the wave vector and $\omega$ the frequency.
The effect of the plane-wave assumption is simply to replace $\partial/\partial
t$ by $-i \omega$ and $\nabla$ by $i \hspace{-1mm} \stackrel{\rightarrow}{k}$.
Form equations (\ref{eq:4}) and (\ref{eq:5}) with the plane-wave solutions, 
Pek\"unl\"u et al. (2002) obtained the dispersion relation
\begin{eqnarray}
\nu \eta k^4 + \Big[v_A^2  - i \omega (\nu + \eta)\Big] k^2 - \omega^2 = 0
\label{eq:29}
\end{eqnarray}

\noindent
This dispersion relation has been used by Dwivedi and Srivastava (2006) for 
propagation and dissipation of Alfv\'{e}n waves in coronal holes. Recently
SD tried to extend this work in an erroneous manner,
as discussed in the following section.

\section{Flaw in the work of SD}                                                                              
In the name of a oblique propagation, SD modified equations (\ref{eq:8}) to 
the form
 \begin{eqnarray}
\stackrel{\rightarrow}{v}_1 = \stackrel{\rightarrow}{v} \ \mbox{e}^{i(\stackrel
{\rightarrow}{k} . \stackrel{\rightarrow}{r} |\cos \theta| - \omega t)} 
\hspace{3cm} \stackrel{\rightarrow}{B}_1 = \stackrel{\rightarrow}{B} \ 
\mbox{e}^{i(\stackrel{\rightarrow}{k} . \stackrel{\rightarrow}{r} |\cos 
\theta| - \omega t)} \label{eq:9}
\end{eqnarray}

\noindent
In the paper of SD, it is not clear where this angle $\theta$ exists; it is 
obviously not between $\stackrel{\rightarrow}{k}$ and $\stackrel{\rightarrow}
{r}$. To our 
understanding, these expressions (\ref{eq:9}) have no physical meaning.
By doing so, SD replaced $\partial/\partial t$ by $-i \omega$ and $\nabla$ by 
$i \hspace{-1mm} \stackrel{\rightarrow}{k} |\cos \theta|$ in the equations 
(\ref{eq:4}) and (\ref{eq:5}), and obtained the dispersion relation 
\begin{eqnarray}
\nu \eta \cos^4 \theta \ k^4  + \Big[v_A^2  - i \omega (\nu + \eta) \Big] 
\cos^2 \theta k^2 - \omega^2 =  0 \label{eq:30}
\end{eqnarray}

\noindent
After using this dispersion relation, SD calculated
various parameters. 

Earlier also, an unphysical situation was created by the same group. In their 
paper, Dwivedi and Pandey (2003, henceforth DP) considered a new energy equation and concluded 
that the results of Porter et al. (1994) and of Pek\"unl\"u et al. (2002) were 
in error and advised to replace the energy equation by Equation (4) of  DP. 
As discussed by Klimchuk et al. (2004) also, the energy equation of DP is 
wrong from the dimension point of view. Thus, the advice of DP was based on an
unphysical situation. For the paper of DP, the Editor of 
Solar Physics invited DP to prepare an apology.

Our objection is that the dispersion relation (\ref{eq:30}) derived by 
using the equations (\ref{eq:9}) is not correct. To our knowledge, no waves can be 
expressed by the relations (\ref{eq:9}). In view of this objection, the investigation of SD is not reliable.

\section{Acknowledgments}

This work was done during the visit of the authors to the IUCAA, Pune. Our 
special thanks are due to Prof. A.K. Kembhavi for encouragement.


\begin{thebibliography}{10}


\bibitem{} Dwivedi, B.N. and Srivastava, A.K.: 2006, {\it Solar Phys.} 
{\bf 237}, 143.

\bibitem{} Dwivedi, B.N. and Pandey, V.S.: 2003, {\it Solar Phys.} {\bf 216}, 
59.


\bibitem{} Klimchuk, J.A., Porter, L.J. and Sturrock, P.A.: 2004, {\it Solar 
Phys.} {\bf 221}, 47.

\bibitem{} Kumar, N., Kumar, P. and Singh, S.: 2006, {\it Astron. Astrophs.} 
{\bf 453}, 1067.

\bibitem{} Pek\"unl\"u, E.R., Bozkurt, Z., Afsar, M., Soydugan, E. and 
Soydugan, F.: 2002 , {\it Mon. Notices Roy. Astron. Soc.} {\bf 336}, 1195.

\bibitem{} Porter, L.J., Klimchuk, J.A. and Sturrock, P.A.: 1994, {\it Astrophs.
J.} {\bf 435}, 482.

 
\bibitem{} Priest, E.R.: 1982, {\it Solar Magnetohydrodynamics}, D. Reidel 
Publishing Company, Dordrecht, Holland. 

\bibitem{} Srivastava, A.K. and  Dwivedi, B.N.: 2007, {\it J. Astrophs. Astr.} 
{\bf 28}, 1.
\end{thebibliography}
\end{document}